\documentclass{iaus}
\usepackage{graphicx}
\usepackage{amssymb}

\title[IAU 275.~~Radio quiet black hole binaries] 
{On the nature of the ``radio quiet''\\ black hole binaries}

\author[Paolo Soleri \& Rob Fender]   
{Paolo Soleri$^1$
 \and Rob Fender$^2$}

\affiliation{$^1$Kapteyn Astronomical Institute, University of Groningen, \\ PO BOX 800,
9700 AV Groningen, the Netherlands \\ email: {\tt soleri@astro.rug.nl} \\[\affilskip]
$^2$School of Physics and Astronomy, University of Southampton, \\ Hampshire SO17 1BJ
Southampton, United Kingdom \\email: {\tt R.Fender@soton.ac.uk}}

\pubyear{2010}
\volume{275}  
\pagerange{1--4}
\jname{Jets at all scales}
\editors{G.E. Romero, R.A. Sunyaev \& T. Belloni, eds.}
\begin{document}

\maketitle

\begin{abstract}
The accretion/ejection coupling in accreting black hole binaries has been described by 
empirical relations between the X-ray/radio and X-ray/optical-infrared luminosities.
These correlations were initially supposed to be universal.
However, recently many sources have been found to produce jets that, given certain accretion-powered
luminosities, are fainter than expected from the correlations.
This shows that black holes with similar accretion flows can produce a broad range of outflows in power
Here we discuss whether typical parameters of the binary system, as well as the properties of the
outburst, produce any effect on the energy output in the jet.
We also define a jet-toy model in which the bulk Lorentz factor becomes larger than $\sim 1$ above $\sim 0.1\%$ of the
Eddington luminosity. We finally compare the ``radio quiet'' black holes with the neutron stars.
\end{abstract}

\firstsection 
\section{Introduction}
Relativistic ejections (jets) are a common consequence of accretion processes onto stellar-mass black holes.
In the low/hard state (LHS) and in the quiescent state of black hole candidates (BHCs) a
compact, steady jet is on.
The jet is highly quenched in the high/soft state (HSS) of BHCs (see \cite[Fender 2010]{Fender2010} for a review).\\
\cite[Corbel et al. (2003)]{Corbel2003} and \cite[Gallo et al. (2003)]{GFP2003} found that the radio luminosity
of many BHCs in the LHS correlates over
several orders of magnitude with the X-ray luminosity. They proposed that a correlation of the form $L_X \propto L_{R}^{0.58\pm0.16}$
(where $L_X$ and $L_R$ are the X-ray and radio luminosities) could be universal and
also valid for sources in quiescence (\cite[Gallo et al. 2006]{Gallo2006}). This relation describes a coupling between the accretion
processes and the ejection mechanisms.
A similar correlation also hold between the the X-ray and the optical/infrared (IR) luminosities (Russell et al. 2006).\\
However, in the past few years, several ``radio quiet'' outliers have been found (\cite[Xue \& Cui 2007]{Xue_Cui2007};
\cite[Gallo 2007]{Gallo2007}).
These sources seem to feature similar X-ray luminosities to other
BHCs but are characterized by a radio emission that, given a certain X-ray luminosity, is fainter than
expected from the radio/X-ray correlation.
It is possible that a correlation with similar slope but lower normalization than the other BHCs could describe this
discrepancy, at least in a few sources (e.g. \cite[Soleri et al. 2010]{Soleri2010}). 
If confirmed, this would suggest that some other parameters might be tuning the accretion-ejection coupling, allowing accretion
flows with similar radiative efficiency to produce a broad range of outflows.\\
\cite[Casella \& Pe'er (2009)]{PG_Asaf09} suggested that different values of the jet magnetic field can cause a quenching of the
radio emission, without influencing the energy output in the X rays.
\cite[Fender et al. (2010)]{FGR2010} showed that, if our measures of the spin and the estimates of the jet power are correct,
the spin does not play any role in powering jets from BHCs.

In this conference Proceedings we investigate whether there is a connection between the values of some binary parameters and
properties of the outburst of 17 BHCs (listed in Figure \ref{fig:toy_model}, left-hand panel) and the compact
steady-jet power.
We will follow the approach presented in \cite[Fender et al. (2010)]{Fender2010} to use the normalizations of the radio/X-ray and
IR/X-ray correlations as a proxy for the jet power.
The data used to calculate the normalizations are from \cite[Gallo et al. (2003)]{GFP2003},
\cite[Gallo et al. (2006)]{Gallo2006}, \cite[Gallo (2007)]{Gallo2007},
\cite[Russell et al. (2006)]{Russell2006}, \cite[Russell et al. (2007)]{Russell2007} and
\cite[Soleri et al. (2010)]{Soleri2010}. We develop a jet-toy model to study whether
de-boosting effects can explain the scatter around the radio/X-ray correlation.
We also compare the ``radio quiet'' BHCs to the accreting neutron star (NS) X-ray binaries.

\section{BHC properties and jet power}
\begin{figure}[tb]
\begin{center}
 \includegraphics[width=7.5cm]{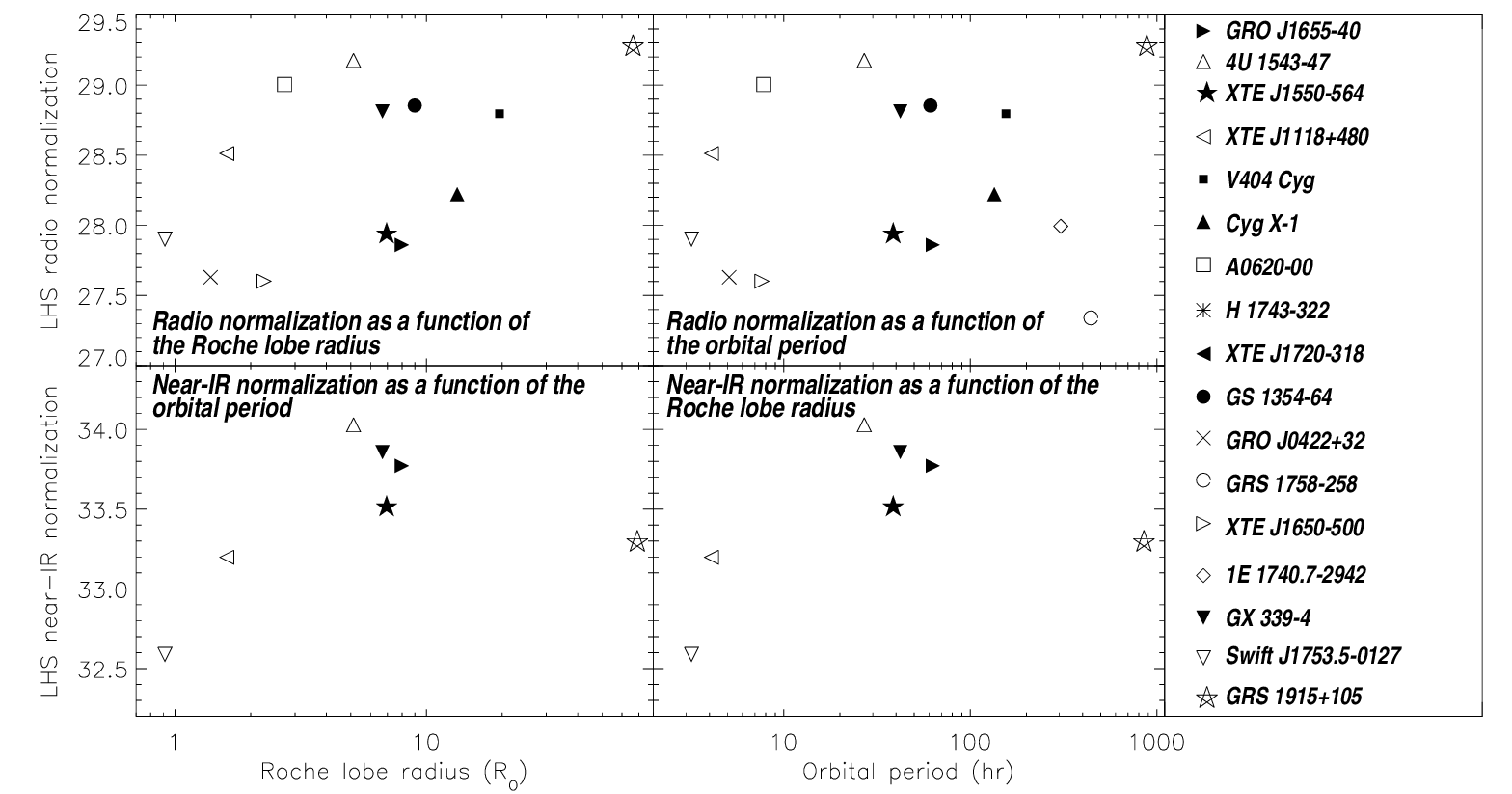}
 \includegraphics[width=5.5cm]{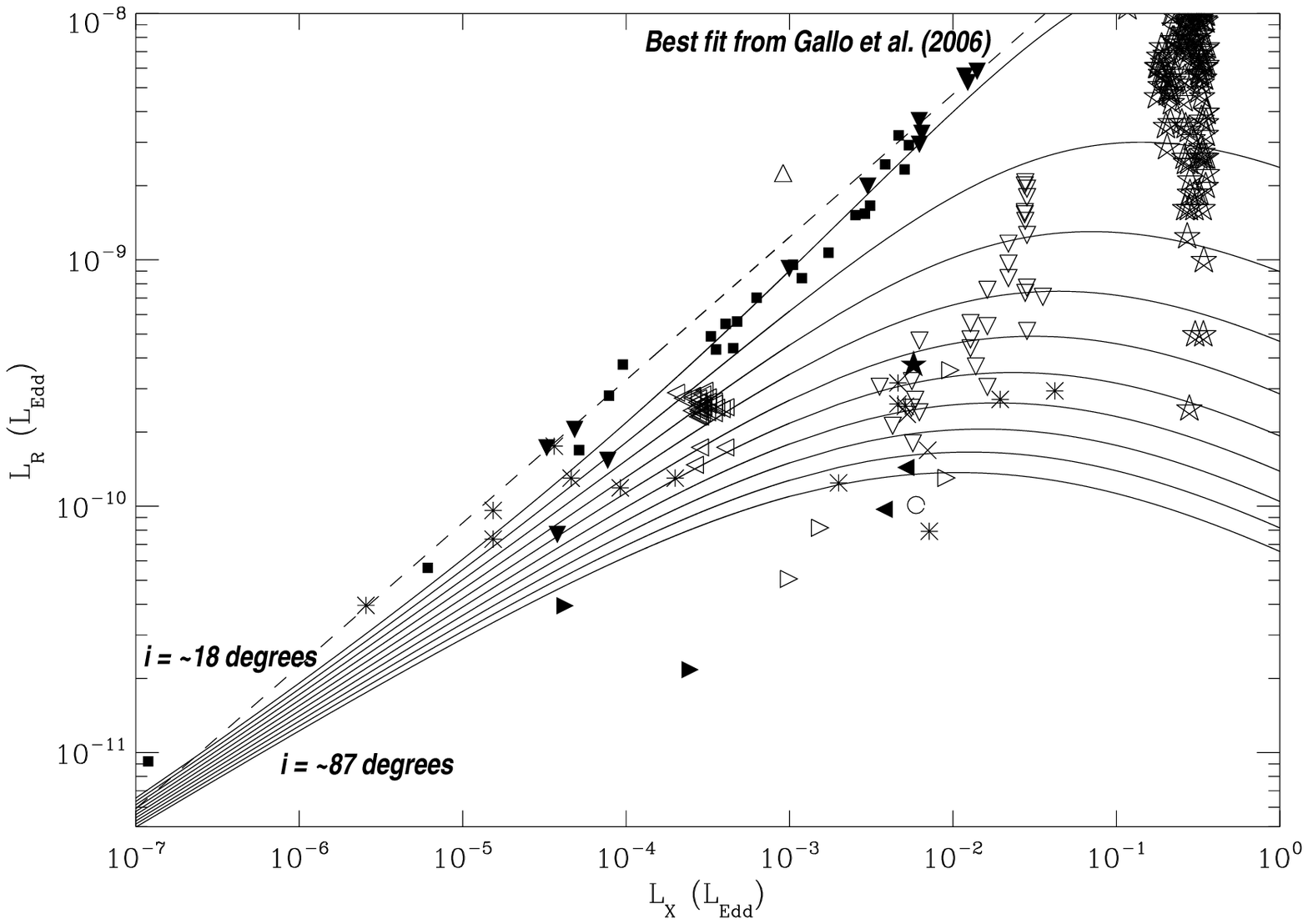} 
 \caption{{\it Left-hand panel:} Radio and near-IR normalizations as a function of the orbital period
and the size of the Roche lobe of the BHCs. Our BHC sample is listed in the inset. The inset also shows a key to the symbols.
{\it Right-hand panel:} Values of the radio luminosity expected from our toy model for 10 viewing angles, in Eddington units.
See the left-hand panel for a key to the symbols.}
   \label{fig:toy_model}
\end{center}
\end{figure}

Since the accretion disc occupies $\sim 70\%$ of the Roche lobe of the black hole, we calculated the size of the Roche lobe of the accretor as a measure of the disc size.
Figure \ref{fig:toy_model} (left-hand panel) shows the radio and near-IR normalizations as a function of the size of
the Roche lobe of the black hole and the orbital period of the binary.
To test whether there is any correlation between the jet power and these two orbital parameters, we calculated the Spearman rank correlation coefficients. The values of the correlation coefficients
$\rho$, as well as the null hypothesis probabilities (the probability that the data are not correlated), are reported in
Table \ref{tab:log_spearman}. Clearly no correlation is present.

We also investigate the dependence of the radio and near-IR normalizations on the inclination between the jet axis
and the line of sight. We will refer to this angle $i$ as either inclination or viewing angle.
Casella et al (2010) recently showed that compact-steady jets from BHCs can have rather high bulk Lorentz factor $\Gamma >2$.
This result suggests that de-boosting effects can become important, not only at high viewing angles. In our analysis
we are assuming that the X-ray emission is un-beamed (but see Fender 2010 and references therein).
To test if a correlation exists, we calculated the Spearman coefficients $\rho$. We show them in Table \ref{tab:log_spearman}.
In the case of the near-IR normalizations, we obtained $\rho \sim -0.9$, with a probability for the null hypothesis
of $\sim 2 \%$. This suggests that there is an anticorrelation between the inclination angle and the near-IR normalization.
However, the lack of data points (only 7) might have biased this result.

During an outburst, BHCs usually show a transition to the HSS. However, some sources
spend the whole outburst in the LHS (or in the LHS and in the intermediate states), without transiting to the soft states
(see \cite[Belloni 2009]{Belloni2009} and references therein).
We investigated whether the type of outburst (LHS only or with a transition to the soft states) affects the jet power, but we could not find any obvious dependence.

\section{A jet-toy model}
Here we define a jet-toy model. The aim is to test whether a dependence of the bulk Lorentz factor of the
jet $\Gamma$ on the accretion powered X-ray luminosity might qualitatively describe the scatter around
the radio/X-ray correlation and the ``radio quiet'' BHCs population.
We will consider a $\Gamma$ Lorentz factor that becomes larger than $\sim 1.4$ above $\sim 0.1 \%$ of the Eddington
luminosity $L_{Edd}$. This assumption is based on the fact that compact
steady jets are usually thought to be mildly relativistic ($\Gamma \leq 2$ but see \cite[Casella et al. 2010]{Casella2010})
in the LHS while major relativistic ejections ($\Gamma \geq 2$) are tentatively associated with the transition from the hard to
the soft states. We considered a uniform distribution of 10 values of $\mbox{cos}\, i$ between 0 and 1.
Figure \ref{fig:toy_model} (right-hand panel) illustrates the results from our toy model: it clearly results in a distribution in the ($L_X,L_R$)
plane which broadens at higher luminosities. The toy model can qualitatively reproduce the scatter around the radio/X-ray correlation.
\begin{table}
  \begin{center}
  \caption{Spearman rank correlation test for our data points. The number of data points, as well as the
probabilities for the null hypothesis are reported.}
  \label{tab:log_spearman}
 {\scriptsize
  \begin{tabular}{l c c c}
\hline  
Normalization          & Number of points  & Spearman coefficient $\rho$   & Probability for the null hypothesis (\%)   \\
                                       \multicolumn{4}{c}{{\bf Size of the Roche lobe}}                               \\                     
radio                  &       13        &          0.5                  &                 11.0                       \\
near IR                &       7         &          0.3                  &                 43.0                       \\
                              \multicolumn{4}{c}{{\bf Orbital period}}                                                      \\
radio                  &      15         &          0.2                  &                 54.2                       \\
near IR                &      7          &          0.4                  &                 33.2                       \\
                              \multicolumn{4}{c}{{\bf Inclination angle}}                                                         \\
radio                  &       13        &          -0.2                 &                 44.7                       \\
near IR                &       7         &          -0.9                 &                 2.4                        \\
\hline 
  \end{tabular}
  }
 \end{center}
\end{table}

\section{Comparison with neutron stars}
NSs are known to be fainter in radio than BHCs, given a certain X-ray luminosity, by a factor $\gtrsim 30$
(see e.g. \cite[Migliari \& Fender 2006]{MigliariFender2006}). This difference in radio power can be reduced to
a factor $\gtrsim 7$ if a mass correction from the fundamental plane of black hole activity
(\cite[Merloni et al. 2003]{Merloni2003}) is applied.
We will now compare the ``radio quiet'' BHCs to the population of NSs that have been detected in radio.
Our sample of NSs includes the same data points as in \cite[Migliari \& Fender (2006)]{MigliariFender2006}
with the addition of points from recent observations of Aql~X-1, 4U~0614-091 and IGR~J00291+5934
(\cite[Tudose et al. 2009]{Tudose2009}, \cite[Migliari et al. 2010]{Migliari2010} and
\cite[Lewis et al. 2010]{Lewis2010}, respectively).
To test whether the ``radio quiet'' BHCs and the NSs are statistically distinguishable in the ($L_X,L_R$) plane,
we performed a two-dimensional Kolmogorov-Smirnov (K-S) test.
The K-S test shows that the probability that the ``radio quiet'' BHCs and the NSs are statistically indistinguishable
(i.e. the probability of the null hypothesis) is different from 0, despite being small ($P \sim 0.13 \%$),
if a mass correction is applied. If we do not apply a mass correction, the probability of the null hypothesis
is consistent with 0: the two groups constitute two different populations.

\section{Conclusions}
We examined three characteristic parameters of BHCs and the properties of their outbursts to test whether they
regulate the energy output in the jet.
If our estimates of the jet power are correct, both the orbital period and the size of the accretion disc are
not related to the radio and near-IR jet power.
We could also not find any association between the jet power and the type of outburst (with or without a transition to the HSS).
We did not find any association between the
viewing angles and the jet power inferred from radio observations. The jet power obtained from near-IR measurements
decreases when the inclination angle increases. This result could favour a scenario in which the jet
decelerates moving from the IR-emitting to the radio-emitting part.\\
We defined a jet-toy model in which the jet Lorentz factor becomes larger than $\sim 1$ above $0.1 \% \, L_{Edd}$.
The model results in a distribution in the ($L_X,L_R$) plane which broadens at high luminosities.
However, the model has several limitations, for instance it can not reproduce the measured inclination angles
of the BHCs in our sample.\\
We finally compared the ``radio quiet'' BHCs to the NSs. A two-dimensional K-S test can not completely
rule out the possibility that the two families are statistically indistinguishable in the ($L_X,L_R$) plane, if a mass
correction is applied. This result suggests that some ``radio quiet'' BHCs could actually be NSs; alternatively it suggests
that some BHCs feature a disc-jet coupling similar to NSs.

\end{document}